\documentstyle[aps,preprint,tighten]{revtex}

\begin{document}
%\draft


\preprint{\vbox{Submitted to Physics Letters B\hfill SNUTP 94--125\\
                                     \null\hfill UM 95--067\\
                                     \null\hfill DOE/ER/40762--053}}

\title{QCD sum rules and chiral logarithms}
\author{Su Houng Lee$^{\rm a,}$\thanks{E-mail: suhoung@phya.yonsei.ac.kr},
Seungho Choe$^{\rm a,}$\thanks{E-mail: schoe@bubble.yonsei.ac.kr},
Thomas D. Cohen$^{\rm b,}$\thanks{E-mail: cohen@quark.umd.edu}$^{\rm ,}$%
\thanks{Permanent address:
Department of Physics and Center for Theoretical Physics,
University of Maryland, College Park, MD 20742, USA},
David K. Griegel$^{\rm c,}$\thanks{E-mail: griegel@quark.umd.edu}}
\address{$^{\rm a}$Department of Physics\\
Yonsei University, Seoul 120--749, Korea}
\address{$^{\rm b}$Department of Physics and Institute for Nuclear Theory\\
University of Washington, Seattle, WA 98195, USA}
\address{$^{\rm c}$Department of Physics and Center for Theoretical Physics\\
University of Maryland, College Park, MD 20742, USA}
\date{November 29, 1994}
\maketitle
\begin{abstract}
Standard QCD sum-rule analyses of the nucleon mass
give results that are inconsistent with
chiral perturbation theory due to an overly simple continuum ansatz on
the phenomenological side of the sum rule.
We show that a careful treatment of the continuum, including $\pi$-$N$
states and other states with virtual pions, resolves the inconsistency
associated with chiral logs.
\end{abstract}
\pacs{PACS numbers: ???}


While we believe that QCD is the theory underlying strong interactions,
the problem of describing low-energy hadronic physics remains an active
and controversial field.
The essential difficulty is that QCD remains intractable in this regime;
thus progress has been made primarily through partial
treatments of the problem, where knowledge of QCD is supplemented by
phenomenological input.
Chiral perturbation theory \cite{gasser1,gasser2} and QCD sum rules
\cite{shifman,reinders} are two largely orthogonal approaches of this
variety that have played a significant role in describing and
explaining low-energy hadronic phenomena.
In this letter, we explore the relationship between the two
approaches.

It is generally believed that chiral perturbation theory accurately
describes all low-energy observables of QCD in the limit of light
current quark masses \cite{gasser1,gasser2}.
In the present context, however, we will only be interested in the
leading nonanalytic behavior of observables as a function of $m_q$, the
average of the up and down current quark masses.
It should be noted that this nonanalytic behavior can be determined
without the full machinery of chiral perturbation theory---it depends
only on the existence of dispersion relations and a
pseudo-Goldstone pion.

The basic idea of QCD sum rules \cite{shifman,reinders} is to
extrapolate from the large spacelike momentum region, where we know how
to treat QCD, down to the low timelike region relevant for hadronic
physics.
This requires three steps.
First, a time-ordered correlation function of interpolating fields is
calculated using the operator product expansion (OPE), which gives a
large-momentum expression for the correlator; information about the
nonperturbative QCD vacuum enters this description via nonvanishing
condensates.
Second, the correlator is related to the spectral density using a
dispersion relation; a model with a small number of parameters is then
made for the spectral density, which gives a phenomenological
description of the correlator.
Finally, the OPE and phenomenological sides are matched in a manner
based on the analytic properties of the correlator and asymptotic
freedom, and the spectral parameters, such as masses and couplings, are
extracted.

In principle, an exact evaluation of the correlator in
terms of QCD degrees of freedom must reproduce all of the low-energy
hadronic physics, including the physics associated with chiral
perturbation theory.
However, all practical implementations of QCD sum rules are not exact,
and it is not guaranteed that the relations thus obtained are
consistent with low-energy constraints.

In a previous work \cite{griegel}, two of us (DKG and TDC) pointed out
inconsistencies in the leading nonanalytic behavior in $m_q$ between
the usual QCD sum-rule treatment of the nucleon mass (based on a simple
continuum ansatz in the spectral density) and the chiral perturbation
theory description.
The simplest QCD sum-rule formula for the nucleon mass \cite{ioffe1}
directly relates the nucleon mass to the quark condensate.
The inconsistency originates from the fact that the quark condensate
has a leading nonanalytic term proportional to $m_q\ln m_q$
\cite{langacher,novikov,gasser1}, whereas the nucleon mass is known to
have a leading nonanalytic term proportional to $m_q^{3/2}$, but no
$m_q\ln m_q$ term \cite{gasser2}.
In Ref.~\cite{griegel}, it was argued that the origin of this problem
clearly lies in the overly simple model for the continuum usually used
in QCD sum rules.
The consequence of this was nontrivial---a rough estimate of the
uncertainty due to the inconsistent nonanalytic behavior was of order
$100\,\text{MeV}$ for the nucleon mass, with a
comparable uncertainty for the $\sigma$ term.

In this letter we will show that the inconsistencies between chiral
perturbation theory and QCD sum rules disappear if the physics of virtual
pions is properly taken into account on the phenomenological side of the sum
rule.
This is quite reminiscent of the case of octet baryons at
finite temperature, as described in Ref.~\cite{koike}.
The consequence of this for QCD sum-rule descriptions of the nucleon
will be discussed at the conclusion of this letter.


QCD sum-rule analyses of the nucleon mass \cite{ioffe1} are based on
the time-ordered correlation function $\Pi_N(q)$ defined by
\begin{equation}
\Pi_N(q)\equiv i\int d^4x\, e^{iq\cdot x}
\langle{\rm vac}|T\eta_N(x)\overline{\eta}_N(0)|{\rm vac}\rangle\ ,
\end{equation}
where $|{\rm vac}\rangle$ is the physical nonperturbative vacuum state,
and $\eta_N$ is an interpolating field with the spin and isospin of a
nucleon, but with indefinite parity.
We use the usual Ioffe interpolating fields for the proton and
neutron \cite{ioffe1,ioffe2}:
\begin{equation}
\eta_p=\epsilon_{abc}(u_a^T C\gamma_\mu u_b)\gamma_5\gamma^\mu d_c\ ,\ \
\eta_n=-\epsilon_{abc}(d_a^T C\gamma_\mu d_b)\gamma_5\gamma^\mu u_c\ ,
\label{int_field}
\end{equation}
where $u_a$ and $d_a$ are up and down quark fields ($a$ is a color index),
$T$ denotes a transpose in Dirac space,
and $C$ is the charge-conjugation matrix.
These choices of the interpolating fields are not unique;
one can also choose
\begin{equation}
\eta_p^\prime=\epsilon_{abc}
(u_a^T C\sigma_{\mu\nu}u_b)\gamma_5\sigma^{\mu\nu}d_c\ ,\ \
\eta_n^\prime=-\epsilon_{abc}
(d_a^T C\sigma_{\mu\nu}d_b)\gamma_5\sigma^{\mu\nu}u_c\ ,
\end{equation}
but $\eta_p$ and $\eta_n$ are the standard---and most
effective---interpolating fields for QCD sum-rule studies of the nucleon
\cite{ioffe1,ioffe2}.
Our conclusions will not, however, depend on this choice.

Lorentz covariance and parity invariance imply that the Dirac structure
of $\Pi_N(q)$ is of the form \cite{bjorken,itzykson}
\begin{equation}
\Pi_N(q)\equiv\Pi_1(q^2)+\Pi_q(q^2)\rlap{/}{q}\ .
\label{corr_decomp}
\end{equation}
Asymptotic freedom and analyticity imply that the Lorentz scalar function
$\Pi_i(s)$ ($i=\{1,q\}$) satisfies the sum rule
\cite{fischer,gimenez,griegel}
\begin{equation}
\int_0^\infty ds\,W_i(s)\Delta\Pi_i(s)
=\int_0^\infty ds\,W_i(s)\Delta\Pi_i^{\rm OPE}(s)\ ,
\end{equation}
where the discontinuity,
\begin{equation}
\Delta\Pi_i(s)\equiv\lim_{\epsilon\rightarrow 0^+}
\Pi_i(s+i\epsilon)-\Pi_i(s-i\epsilon)\ ,
\label{sumrule}
\end{equation}
is proportional to the spectral density, $\Pi_i^{\rm OPE}(s)$
denotes the scalar function evaluated using the OPE,
and the weighting function $W_i(s)$ is an arbitrary entire function.
The weighting function is chosen so as to improve the convergence of
the OPE while simultaneously strongly weighting the nucleon pole
contribution to the sum rule relative to the continuum contribution.
The usual QCD sum rules based on the Borel transform \cite{shifman,reinders}
can be obtained by
choosing $W_1(s)=W_q(s)=e^{-s/M^2}$, where $M$ is
known as the Borel mass.

The correlator contains contributions from the nucleon pole and from
higher-mass (continuum) states with the quantum numbers of a nucleon;
thus the correlator can be parameterized as \cite{bjorken,itzykson}
\begin{equation}
\Pi_N(q)=-\lambda_N^2{1\over\rlap{/}{q}-M_N}+\Pi_N^{\rm cont}(q)\ ,
\label{pi_phen}
\end{equation}
where $\lambda_N$ specifies the strength of the coupling between the
interpolating field and the physical nucleon state
[see Eq.~(\ref{lambda_def})],
and $\Pi_N^{\rm cont}(q)$ denotes the contribution from the continuum.

On the other hand, the OPE expresses the correlator in terms of vacuum
condensates.
With the interpolating fields in Eq.~(\ref{int_field}),
one obtains \cite{ioffe1,leinweber}
\begin{eqnarray}
\Pi_1^{\rm OPE}(s)&=&
{s\over 4\pi^2}\ln(-s)\langle\overline{q}q\rangle
-{11\over 288\pi^2 s}\langle\overline{q}q\rangle\langle(gG)^2\rangle
+\cdots\ ,
\\*
\Pi_q^{\rm OPE}(s)&=&-{s^2\over 64\pi^4}\ln(-s)
-{1\over 128\pi^4}\ln(-s)\langle(gG)^2\rangle
-{2\over 3s}\langle\overline{q}q\rangle^2+\cdots\ ,
\end{eqnarray}
where $\langle\overline{q}q\rangle\simeq -(225\,\text{MeV})^3$
[$\overline{q}q\equiv{1\over 2}(\overline{u}u+\overline{d}d)$]
and $\langle(gG)^2\rangle\simeq 0.5\,\text{GeV}^4$ are the quark and
gluon condensates \cite{shifman,reinders}.

The usual strategy for extracting the nucleon mass in QCD sum rules is
to separate the nucleon pole contributions to the sum rules from the
OPE and continuum contributions.
For simplicity, we use the same weighting function $W(s)$ in both sum rules.
The nucleon mass is then obtained by taking the ratio of the sum rule
for $\Pi_1$ to the sum rule for $\Pi_q$ \cite{ioffe1}:
\begin{equation}
M_N={\displaystyle{\int_0^\infty dt\,W(t)
\left[\Delta\Pi_1^{\rm OPE}(t)-\Delta\Pi_1^{\rm cont}(t)\right]}
\over
\displaystyle{\int_0^\infty ds\,W(s)
\left[\Delta\Pi_q^{\rm OPE}(s)-\Delta\Pi_q^{\rm cont}(s)\right]}}\ .
\label{mnucsr}
\end{equation}

The formula for the nucleon mass in Eq.~(\ref{mnucsr}) has a potential
problem when approaching the chiral limit \cite{griegel}.
As we will show below, $\Delta\Pi_1^{\rm OPE}$ contains an
$O(m_q\ln m_q)$ chiral log, whereas $\Delta\Pi_q^{\rm OPE}$ and the
nucleon mass \cite{gasser2} lack this nonanalytic term near the chiral
limit.
This result is somewhat more general than that of Ref.~\cite{griegel},
where it was pointed out that the {\it leading\/} term in
$\Delta\Pi_1^{\rm OPE}$, which is proportional to
$\langle\overline{q}q\rangle$, has a well-known
\cite{langacher,novikov,gasser1} $O(m_q\ln m_q)$ chiral log.

The physics associated with the chiral logs can be thought of as
arising from virtual pions.
As long as one studies the correlation function for $q^2$ sufficiently
far from the nucleon pole ($|q^2-M_N^2|\gg 2M_N m_\pi$), one can use
soft-pion theorems and phase-space arguments to show
\begin{eqnarray}
\Pi_1^{\rm OPE}(q^2)&=&(1-2\epsilon)\overcirc{\Pi}_1^{\rm OPE}(q^2)\ ,
\label{result1}
\\*
\Pi_q^{\rm OPE}(q^2)&=&[1+O(\overcirc{m}_\pi^2)]
\overcirc{\Pi}_q^{\rm OPE}(q^2)\ ,
\label{result2}
\end{eqnarray}
where we have defined
$\overcirc{m}_\pi^2\equiv-2m_q\overcirc{\langle\overline{q}q\rangle}
/\overcirc{f}_\pi^2$.
All other quantities of the form $\overcirc{x}$ denote the value of $x$
in the $m_q=0$ limit.
We have also defined
\begin{equation}
\epsilon\equiv{3\over 64\pi^2\overcirc{f}_\pi^2}
\overcirc{m}_\pi^2\ln{\overcirc{m}_\pi^2\over M_0^2}\ ,
\end{equation}
where $M_0$ is an arbitrary constant.
Changes in $M_0$ can always be absorbed into changes in an analytic
term proportional to $\overcirc{m}_\pi^2$; thus the substitution
$\epsilon\rightarrow\epsilon+O(\overcirc{m}_\pi^2)$ is implicit.
Equivalently, for any weighting function $W(s)$ that
has substantial strength over a region in $s$ that goes well above the
$\pi$-$N$ threshold
\begin{eqnarray}
\int_0^\infty ds\,W(s)\Delta\Pi_1^{\rm OPE}(s)
&=&\int_0^\infty ds\,W(s)(1-2\epsilon)
\Delta\overcirc{\Pi}_1^{\rm OPE}(s)\ ,
\label{chilog}
\\*
\int_0^\infty ds\,W(s)\Delta\Pi_q^{\rm OPE}(s)
&=&\int_0^\infty ds\,W(s)[1+O(\overcirc{m}_\pi^2)]
\Delta\overcirc{\Pi}_q^{\rm OPE}(s)\ .
\label{nochilog}
\end{eqnarray}
Note the restriction in Eqs.~(\ref{result1}) and (\ref{result2}) to
$q^2$ far from the nucleon pole translates to a restriction in
Eqs.~(\ref{chilog}) and (\ref{nochilog}) to weighting functions that
cover a large range in $s$.

In contrast, the nucleon mass is known to have no
$O(\overcirc{m}_\pi^2\ln\overcirc{m}_\pi^2)$ contributions.
Near the chiral limit ($m_\pi\ll M_\Delta-M_N$), the nucleon mass is given by
\begin{equation}
M_N=\overcirc{M}_N+A\overcirc{m}_\pi^2
-{3\overcirc{g}_A^2\over 32\pi\overcirc{f}_\pi^2}\overcirc{m}_\pi^3+\cdots\ ,
\label{mnuc_exp}
\end{equation}
where $A$ is an unknown constant \cite{gasser2}.
The essence of the difficulty is the following:
Consider Eq.~(\ref{mnucsr}) with different values of the quark mass.
The right-hand side of Eq.~(\ref{mnucsr}) has contributions to $M_N$ of
order $m_q\ln m_q$ coming from the quark-mass dependence of the OPE.
On the other hand, we know that $M_N$ has no term that goes as
$m_q\ln m_q$.

In principle, there is no difficulty in reconciling the QCD sum-rule
expression in Eq.~(\ref{mnucsr}) with the results of chiral
perturbation theory.
The continuum contribution in the numerator, for example, can have a
nontrivial chiral behavior that precisely cancels the $m_q\ln m_q$
behavior.
The difficulty is that, in most practical QCD sum-rule calculations,
the model of the spectral function for the continuum is very
crude:
\begin{equation}
\Delta\Pi_i^{\rm cont}(s)=\theta(s-s_0)\Delta\Pi_i^{\rm OPE}\ .
\label{cont_ansatz}
\end{equation}
The issue is whether there is any natural way for the continuum
threshold $s_0$ in this model to vary with $m_q$ in such a manner as to
cancel the spurious $m_q\ln m_q$ behavior.
We will demonstrate that this is impossible.
As we will show here, however, if a better model for the continuum is
used, one that includes $\pi$-$N$ (and higher-mass) continuum states in
a fashion consistent with chiral symmetry, then one automatically
cancels the problematic $m_q\ln m_q$ behavior.

The simplest derivation of Eqs.~(\ref{result1}) and (\ref{result2}) is
based on a spectral representation of the correlator,%
\footnote{We neglect imaginary infinitesimals in the denominator, since
we are interested in the correlator at spacelike momentum transfers.}
\begin{equation}
\Pi_N(q)=\int_{-\infty}^\infty
d\omega\,{\rho_N(\omega,{\bf q})\over\omega-q_0}\ ,
\label{spectral_rep}
\end{equation}
where the spectral density is defined as
\begin{equation}
\rho_N(q)\equiv(2\pi)^3\sum_{n,m}\langle n|\eta_N(0)|m\rangle
\langle m|\overline{\eta}_N(0)|n\rangle
\delta^4(q-p_m+p_n)\ .
\label{spectral_def}
\end{equation}
The states $|n\rangle$ denote a complete set of energy eigenstates,
which are also eigenstates of momentum due to translation invariance.
The double sum in Eq.~(\ref{spectral_def}) is a convenient way to represent
the contributions from positive- and negative-energy states; however, one
must have either $|n\rangle=|{\rm vac}\rangle$ or
$|m\rangle=|{\rm vac}\rangle$ in each term.
To obtain the leading nonanalytic behavior of $\Pi_N$, we consider $N$ and
$\pi$-$N$ intermediate states.
It is clear that other intermediate states will not contribute to the
leading nonanalytic term.

We first evaluate the nucleon- and antinucleon-pole contributions to the
correlator, which correspond to taking
$|n\rangle=|{\rm vac}\rangle$, $|m\rangle=|N(k)\rangle$ and
$|m\rangle=|{\rm vac}\rangle$, $|n\rangle=|\overline{N}(k)\rangle$,
respectively, in Eq.~(\ref{spectral_def}).
Given the definitions
\begin{equation}
\langle{\rm vac}|\eta_N(0)|N(k)\rangle
\equiv\lambda_N u_N(k)\ ,\ \
\langle{\rm vac}|\overline{\eta}_N(0)|\overline{N}(k)\rangle
\equiv\lambda_N\overline{v}_N(k)\ ,
\label{lambda_def}
\end{equation}
where $u_N(k)$ and $\overline{v}_N(k)$ are nucleon and antinucleon spinors,
one readily obtains the result announced in Eq.~(\ref{pi_phen}),
\begin{equation}
\Pi_N^{\rm pole}(q)=-\lambda_N^2{1\over\rlap{/}{q}-M_N}\ ,
\end{equation}
from Eqs.~(\ref{spectral_rep}) and (\ref{spectral_def}).

Next we consider the contributions to the correlator from $\pi$-$N$
and $\pi$-$\overline{N}$ continuum states.
The corresponding states in Eq.~(\ref{spectral_def}) are of the form
$|n\rangle=|{\rm vac}\rangle$,
$|m\rangle=|\pi(k^\prime)N(k)\rangle$ and
$|m\rangle=|{\rm vac}\rangle$,
$|n\rangle=|\pi(k^\prime)\overline{N}(k)\rangle$.
To evaluate the resulting matrix elements, we exploit the transformation
properties of the nucleon interpolating field under SU(2) axial rotations:
\begin{equation}
[Q_5^a,\zeta_N]
=-\gamma_5{\tau^a\over 2}\zeta_N\ ,\ \
\zeta_N\equiv\left(\begin{array}{c} \eta_p \\ \eta_n \end{array}\right)\ ,
\end{equation}
where $Q_5^a$ is the axial charge.
These transformation properties do not depend on the choice of
interpolating field; the same relations hold for
$\eta_N\rightarrow\eta_N^\prime$.
Given the transformation properties for the interpolating field, one can
derive the soft-pion theorem \cite{donoghue},
\begin{eqnarray}
\langle{\rm vac}|\zeta_N|\pi^a N(k)\rangle
&=&-{i\over f_\pi}\langle{\rm vac}|[Q_5^a,\zeta_N]|N(k)\rangle
+O\left(m_\pi,k_\pi\over\Lambda_{\rm had}\right)
\nonumber\\*
&=&{i\over 2f_\pi}\gamma_5\tau^a\langle{\rm vac}|\zeta_N|N(k)\rangle
+O\left(m_\pi,k_\pi\over\Lambda_{\rm had}\right)\ ,
\label{soft}
\end{eqnarray}
where $\Lambda_{\rm had}$ is a typical hadronic scale assumed to be
$\sim 1\,\text{GeV}$.
For $q^2$ far from the nucleon pole, the soft part of the
$\pi$-$N$ continuum yields the following contribution to $\Pi_N(q)$:
\begin{equation}
\Pi_N^{\text{$\pi$-$N$ cont}}(q)=-\epsilon
\gamma_5\Pi_N^{\rm pole}(q)\gamma_5\ .
\label{pi-N}
\end{equation}
The physical origin of this contribution is clear:  The $\pi$-$N$ states
become nearly degenerate with the $N$ states, and the infrared behavior
gives rise to the chiral log behavior.

The second source of chiral logs is the $\lambda_N^2$
factor in the nucleon-pole contribution to the correlator.
Consider the matrix element $\langle{\rm vac}|\eta_N|N\rangle$.
By treating the finite up and down quark masses as a perturbation, one obtains
\begin{equation}
\langle{\rm vac}|\eta_N(0)|N\rangle
=\langle{\rm vac}|\eta_N(0)|N\rangle_{m_q=0}
+i\int d^4 x\,\langle{\rm vac}|T\delta{\cal L}(x)\eta_N(0)|N\rangle_c\ ,
\end{equation}
where we define $\delta{\cal L}\equiv-m_q(\overline{u}u+\overline{d}d)$.
The leading nonanalytic behavior of the time-ordered product is given
by two-pion intermediate states; the resulting matrix elements
are approximated by taking the soft-pion limit.
This yields
\begin{eqnarray}
\langle{\rm vac}|\eta_N|N\rangle
&=&\langle{\rm vac}|\eta_N|N\rangle_{m_q=0}
\nonumber\\*
& &\null
+{i\over 2}m_q\langle{\rm vac}|\overline{u}u+\overline{d}d|\pi^a\pi^b\rangle
\langle\pi^a\pi^b|\eta_N|N\rangle
\int{d^4 k\over(2\pi)^4}{1\over(k^2-m_\pi^2+i\epsilon)^2}\ ,
\end{eqnarray}
where all fields are evaluated at the origin.
{}From PCAC one obtains
\begin{equation}
m_q\langle{\rm vac}|\overline{u}u+\overline{d}d|\pi^a\pi^b\rangle
=m_\pi^2\delta^{ab}\ ,
\end{equation}
and the matrix elment of $\eta_N$ is obtained by applying the soft-pion
theorem twice:
\begin{equation}
\langle\pi^a\pi^b|\eta_N|N\rangle
=-{\delta^{ab}\over 4f_\pi^2}\langle{\rm vac}|\eta_N|N\rangle
+O\left(m_\pi^2,k_\pi^2\over\Lambda_{\rm had}^2\right)\ .
\end{equation}
One then obtains the result
\begin{equation}
\langle{\rm vac}|\eta_N|N\rangle
=\left(1-{\epsilon\over 2}\right)\langle{\rm vac}|\eta_N|N\rangle_{m_q=0}\ ,
\end{equation}
which implies $\lambda_N=(1-\epsilon/2)\overcirc{\lambda}_N$.
Furthermore, the nucleon-pole contribution to the correlator then behaves as
\begin{equation}
\Pi_N^{\rm pole}(q)=(1-\epsilon)\overcirc{\Pi}_N^{\rm pole}(q)\ .
\label{pole_exp}
\end{equation}

We must also determine the chiral behavior of the continuum contributions
to the correlator.
The effect of $\pi$-$N$ continuum states on $\Pi_N$ has been discussed
above [see Eq.(\ref{pi-N})].
The chiral behavior of the remainder of the continuum can be deduced as
follows:
In the chiral limit, this part of the continuum can be represented to
arbitrary accuracy by a discrete set of poles, $N^\ast$,
$N^{\ast\ast}$, $\ldots$, of positive or negative parity.
Away from the chiral limit, each $N^\ast$ is dressed with pions in the
same way as the nucleon; thus the leading nonanalytic contributions to
this part of the correlator arise from the chiral expansion of
$\lambda_{N^\ast}$ and from $\pi$-$N^\ast$ states.
One then obtains relations analogous to Eqs.~(\ref{pole_exp}) and
(\ref{pi-N}) for both positive- and negative-parity states.
The expansion of the continuum contribution to $\Pi_N$ about its value in
the chiral limit, for $q^2$ far from the nucleon pole, is
thus given by
\begin{equation}
\Pi_N^{\rm cont}(q)=-\epsilon
\gamma_5\overcirc{\Pi}_N^{\rm pole}(q)\gamma_5
+(1-\epsilon)\overcirc{\Pi}_N^{\rm cont}(q)
-\epsilon\gamma_5\overcirc{\Pi}_N^{\rm cont}(q)\gamma_5\ .
\label{cont_exp}
\end{equation}
The first term in the preceding equation arises from the fact that,
while the nucleon pole term is, by construction, not part of the
continuum, the $\pi$-$N$ states that give the chiral log in
Eq.~(\ref{pi-N}) are part of the continuum.

By combining Eqs.~(\ref{pole_exp}) and (\ref{cont_exp}), we can determine
the chiral expansion of the correlator.
Decomposing $\Pi_N(q)$ into scalar functions according to
Eq.~(\ref{corr_decomp}), we obtain
\begin{eqnarray}
\Pi_1(q^2)&=&(1-2\epsilon)\overcirc{\Pi}_1(q^2)\ ,
\\*
\Pi_q(q^2)&=&[1+O(\overcirc{m}_\pi^2)]\overcirc{\Pi}_q(q^2)\ .
\end{eqnarray}
These results can be used at high spacelike momentum transfer, where they
must match the OPE description.
Thus one obtains the results stated in
Eqs.~(\ref{result1}) and (\ref{result2}):
\begin{eqnarray}
\Pi_1^{\rm OPE}(q^2)&=&(1-2\epsilon)\overcirc{\Pi}_1^{\rm OPE}(q^2)\ ,
\label{OPE1}
\\*
\Pi_q^{\rm OPE}(q^2)&=&[1+O(\overcirc{m}_\pi^2)]
\overcirc{\Pi}_q^{\rm OPE}(q^2)\ .
\label{OPEq}
\end{eqnarray}
Since Eq.~(\ref{OPE1}) holds for all spacelike $q^2$, it follows that
{\it all\/} terms in $\Pi_1^{\rm OPE}$ must contain the $(1-2\epsilon)$ factor.
In Ref.~\cite{griegel}, it was demonstrated that the {\it leading\/} term in
$\Pi_1^{\rm OPE}$, which is proportional to
$\langle\overline{q}q\rangle$, behaves this way; here we see that all
of the other terms do too.
Similarly, all terms in $\Pi_q^{\rm OPE}$ must go as
$[1+O(\overcirc{m}_\pi^2)]$.

Let us now use the known chiral expansions of $\Pi_N^{\rm OPE}$ and
$\Pi_N^{\rm cont}$ to see whether there is a chiral-log contribution to
$M_N$ in Eq.~(\ref{mnucsr}).
Inserting the results of Eqs.~(\ref{OPE1}), (\ref{OPEq}), and (\ref{cont_exp})
into Eq.~(\ref{mnucsr}) gives
\begin{equation}
M_N=\overcirc{M}_N+O(\overcirc{m}_\pi^2)\ .
\end{equation}
Thus we see that, if the continuum contribution to the correlator is
treated carefully, the spurious chiral-log contributions to the nucleon
mass, discussed in Ref.~\cite{griegel}, cancel.
Thus the QCD sum-rule prediction is consistent with the chiral perturbation
theory description of the nucleon mass given in Eq.~(\ref{mnuc_exp})
to $O(m_q)$.
However, it lacks the $O(m_q^{3/2})$ contribution, which must arise from
{\it subleading\/} nonanalytic contributions to the nucleon correlator.

While it is clear that the inclusion of virtual pions
automatically removes the spurious chiral logs, it is worth exploring
the question of whether the usual continuum ansatz in
Eq.~(\ref{cont_ansatz}) can accomplish the same results.
This is impossible.
Decomposing the continuum contribution to the correlator given in
Eq.~(\ref{cont_exp}) according to Eq.~(\ref{corr_decomp}), one obtains
\begin{eqnarray}
\Pi_1^{\rm cont}(q^2)&=&-\epsilon\overcirc{\Pi}_1^{\rm pole}(q^2)
+(1-2\epsilon)\overcirc{\Pi}_1^{\rm cont}(q^2)\ ,
\\*
\Pi_q^{\rm cont}(q^2)&=&\epsilon\overcirc{\Pi}_q^{\rm pole}(q^2)
+[1+O(\overcirc{m}_\pi^2)]\overcirc{\Pi}_q^{\rm cont}(q^2)\ ,
\end{eqnarray}
and it is straightforward to see that the form given in
Eq.~(\ref{cont_ansatz}) is inconsistent with these results.
Physically, one can understand this by noting that there is spectral
strength very near the nucleon pole coming from low-momentum $\pi$-$N$
states, which give rise to part of the chiral log.
Alterations in the value of the continuum threshold, however, can only
change the spectral density far above the nucleon pole.

Finally, we consider the implications of our
results for QCD sum rules for the nucleon.
It seems clear that one should include low-lying $\pi$-$N$ continuum
states on the phenomenological side of the sum rule;
this will remove the spurious chiral logs and remove one source of
uncertainty.
The uncertainty from this effect has been estimated to be
$\sim 100\,\text{MeV}$ for both the nucleon mass and $\sigma$ term
\cite{griegel}.
However, we note that, although this removes the rather unsatisfactory
chiral log behavior, it does not completely remove the uncertainties.
In particular, chiral perturbation theory gives a term proportional to
$m_q^{3/2}$ as the leading nonanalytic contribution to the nucleon mass,
which arises from $\pi$-$N$ intermediate states in the nucleon propagator.
We expect this term to originate in a similar manner in a QCD sum-rule
analysis.
It seems certain that the $m_q^{3/2}$ term is lost in the present QCD
sum-rule description due to the use of the soft-pion theorem in
evaluating matrix elements involving $\pi$-$N$ intermediate states
[see Eq.~(\ref{soft})];
presumably, it can be reproduced with an adequate description
of the soft-to-hard pion correction.
In fact, the $m_q^{3/2}$ term {\it is\/} recovered using a simple model
for the soft-to-hard correction based on pseudovector $\pi$-$N$
coupling; however, we know of no satisfactory, model-independent
approach.
The lack of the $m_q^{3/2}$ term leads to uncertainties in
the sum-rule results of $\sim 15\,\text{MeV}$ and $\sim 20\,\text{MeV}$
for the nucleon mass and $\sigma$ term, respectively \cite{griegel}.


\acknowledgements
SHL, TDC, and DKG thank the Institute for Nuclear Theory at the
University of Washington for its hospitality.
SHL and TDC also thank the Department of Physics at the University of
Washington.
The work of SHL and SC was supported by the Basic Science
Research Institute program of the Korean Ministry of Education through Grant
No.\ BSRI--94--2425, by KOSEF through the CTP at Seoul National
University, and by a Yonsei University Research Grant.
The work of TDC and DKG was supported by the U.S. Department of Energy
through Grant No.\ DE--FG02--93ER--40762.
The work of TDC was also supported by the U.S. National Science
Foundation through Grant No.\ PHY--9058487.




\begin{references}
%
\bibitem{gasser1}J. Gasser and H. Leutwyler,
Ann.\ of Phys.\ 158 (1984) 142;
Nucl.\ Phys.\ B250 (1985) 465.
%
\bibitem{gasser2}J. Gasser, M.~E. Sainio, and A. \v{S}varc,
Nucl.\ Phys.\ B307 (1988) 779.
%
\bibitem{shifman} M.~A. Shifman, A.~I. Vainshtein and V.~I. Zakharov,
Nucl.\ Phys.\ B147 (1979) 385;
B147 (1979) 448;
B147 (1979) 519.
%
\bibitem{reinders}L.~J. Reinders, H. Rubinstein, and S. Yazaki,
Phys.\ Rep.\ 127 (1985) 1.
%
\bibitem{griegel}D.~K. Griegel and T. D. Cohen,
Phys.\ Lett.\  B 333 (1994) 27.
%
\bibitem{ioffe1}B.~L. Ioffe,
Nucl.\ Phys.\ B188 (1981) 317;
B191 (1981) 591(E).
%
\bibitem{langacher}P. Langacker and H. Pagels,
Phys.\ Rev.\ D 8 (1973) 4595.
%
\bibitem{novikov}V.~A. Novikov, M.~A. Shifman, A.~I. Vainshtein
and V.~I. Zakharov,
Nucl.\ Phys.\ B191 (1981) 301.
%
\bibitem{koike}Y. Koike,
Phys.\ Rev.\ D 48 (1993) 2313.
%
\bibitem{ioffe2}B.~L. Ioffe,
Z.\ Phys.\ C 18 (1983) 67.
%
\bibitem{bjorken}J.~D. Bjorken and S.~D. Drell,
Relativistic Quantum Fields
(McGraw-Hill, New York, 1965).
%
\bibitem{itzykson}C. Itzykson and J.-B. Zuber,
Quantum Field Theory
(McGraw-Hill, New York, 1980).
%
\bibitem{fischer}J. Fischer and P. Kol\'{a}\v{r},
Z. Phys.\ C 34 (1987) 375.
%
\bibitem{gimenez}V. Gim\'{e}nez, J. Bordes, and J. Pe\~{n}arrocha,
Nucl.\ Phys.\ B357 (1991) 3.
%
\bibitem{leinweber}D. B. Leinweber,
Ann.\ of Phys.\ 198 (1990) 203.
%
\bibitem{donoghue}J.~F. Donoghue, E. Golowich, and B.~R. Holstein,
Dynamics of the Standard Model
(Cambridge University Press, Cambridge, 1992).
%
\end{references}
\end{document}